\documentclass[acmsmall, nonacm]{acmart}
\AtBeginDocument{%
  }

\usepackage{todonotes}
\usepackage{pifont}
\usepackage{xspace}
\usepackage{soul}
\usepackage[normalem]{ulem}
\usepackage{caption}
\usepackage{makecell}
\usepackage{enumitem}
\usepackage{listings}
\usepackage{algorithm}
\usepackage{algpseudocode}
\usepackage{xurl}
\usepackage{wrapfig}

\newcommand\blue[1]{{#1}}
\newcommand\bblue[1]{{#1}}

\usepackage{titlesec}
\usepackage{anyfontsize}

\newcommand{\one}{({\em i}\/)\xspace}
\newcommand{\two}{({\em ii}\/)\xspace}
\newcommand{\three}{({\em iii}\/)\xspace}
\newcommand{\four}{({\em iv}\/)\xspace}

\def\eg{\emph{e.g.,}\xspace}

\def\ie{\emph{i.e.,}\xspace}

\def\vs{\emph{vs.}\xspace}

\newcommand{\pb}[1]{\vspace{0.75ex}\noindent{\bf \em #1}\hspace*{.3em}}

\newcommand\summary[1]{\vspace{0.5ex}\todo[inline,color=white]{\small \spaceskip=0.2em\relax \textbf{Takeaways:} #1}}
\begin{document}

\title{An Empirical Analysis of the Nostr Social Network: Decentralization, Availability, and Replication Overhead}


\author{Yiluo Wei}
\affiliation{%
  \institution{The Hong Kong University of Science and Technology (Guangzhou)}
  \country{China}
}
\email{ywei906@connect.hkust-gz.edu.cn}

\author{Gareth Tyson}
\affiliation{%
  \institution{The Hong Kong University of Science and Technology (Guangzhou)}
  \country{China}
}
\email{gtyson@ust.hk}







\begin{abstract}
    Nostr is a decentralized social network launched in 2022, emphasizing high availability and censorship resistance. Since launching, it has gained substantial attention, boasting over 100 million posts.
    From a user's perspective, it is similar to a micro-blogging service like Twitter.
    However, the underlying systems infrastructure is very different, and Nostr boasts a range of unique features that set it apart.
    Nostr introduces the concept of relays, which act as open storage servers that receive, store, and distribute user posts. Each user is uniquely identified by a public key, ensuring authenticity of posts through digital signatures. Users are able to securely replicate and retrieve posts through multiple relays, which frees them from single-server reliance and enhances post availability, thereby attempting to make Nostr censorship resistant.
    However, this aggressive design also presents challenges, such as the overhead required for extensive post replication and the difficulty in obtaining a global view of post replication locations, which remain unexplored or unaddressed.
    This necessitates a thorough understanding of the Nostr ecosystem; therefore, we conduct the first large-scale study on this topic.
    Our study focuses on two key aspects: Nostr relays and post replication strategies.
    We find that Nostr achieves superior decentralization compared to traditional Fediverse applications. 
    However, relay availability remains a challenge, where financial sustainability (particularly for free-to-use relays) emerges as a contributing factor.
    We also find that the replication of posts across relays enhances censorship-resistance but introduces significant overhead. 
    To address this, we propose two improvements: one to control the number of post replications, and another to reduce the overhead during post retrieval. Via a data-driven evaluation, we demonstrate their ability to reduce overhead without negatively impacting post availability \blue{under the simulated scenarios}.
\end{abstract}

\ccsdesc[500]{Networks~Network measurement}

\keywords{Nostr, Social Network, Decentralization, Measurement, Censorship}

\received{June 2025}
\received[accepted]{September 2025}

\maketitle

\vspace{-1.5ex}
\section{Introduction}
\label{sec:intro}
\vspace{-0.5ex}

Censorship resistance is a major research topic. 
Numerous approaches have been proposed \cite{Khattak2016}, including refraction networking \cite{Bocovich2016}, censorship-resistant CDNs \cite{Zolfaghari2016}, packet manipulators \cite{Bock2019}, and WebRTC-based tools \cite{Barradas2020, 10.1145/3488932.3517419, 298268}. 
Recently, censorship resistance in social media platforms has garnered particular attention \cite{HUANG2024106059, shashlin2023future}.
This has been driven by the monopolistic nature of traditional online social media platforms (such as Facebook and Twitter/X), which has given them significant control over permissible activities, raising growing concerns about censorship and freedom of speech~\cite{facebook-1,free-speech}.

Consequently, there has been a growing movement of technologists and researchers working on decentralized social networks that aim to foster increased transparency, openness, and democracy.  
The ``Fediverse'', composed of server-based federated services, such as Mastodon, rank among the most widely used applications for decentralized social networks \cite{Sigmetrics22-Toxicity}.
These platforms deconstruct their service offerings into independent servers (instances) that anyone can easily setup.
In the simplest scenario, these instances permit users to register and interact locally. However, they also enable cross-instance interaction (\ie federation). This distributed structure offers numerous benefits. For instance, it provides more transparent data ownership, and potentially enhances the overall system's resilience against technical or legal attack.

Despite the benefits, these platforms suffer from key limitations \cite{IMC19-Mastodon, Conext21-Pleroma}. User identities are tied to a single server controlled by an administrator. Thus, instance owners can ban users (just as Twitter/X can), delete content, and even block other servers, thereby imposing censorship.
Perhaps more problematic is that there are no clear incentives for running these servers. Consequently, it can be challenging for part-time enthusiasts to resource such activities.
For example, a recent study found that 11\% of instances are inaccessible for more than half of their lifespan \cite{IMC19-Mastodon}.
This situation puts users at the mercy of a single administrator's personal preferences, which sometimes have been more oppressive than large corporations like Twitter/X \cite{rozenshtein2023moderating}. 

To address the limitations of the Fediverse, a fundamentally different decentralized web platform was launched in 2022, receiving significant financial backing from Jack Dorsey: \emph{Nostr} (Notes and Other Stuff Transmitted by Relays) \cite{nostr}.
From a user's perspective, Nostr exhibits similar functions to a micro-blogging service like Twitter/X, although the actual implementation differs substantially.
Nostr is a decentralized system that is built on a series of open protocols capable of establishing a global social network in a way that detaches users from any specific server.
Publicly, Nostr states that this approach is designed with one key requirement in-mind: attaining censorship resistance against administrators removing content or blocking specific users.

The core innovation is the introduction of \emph{relays}, which operate as open storage servers, loosely akin to the server instances in the Fediverse. Their only responsibility is to receive posts from users, store them, and distribute them to others upon request. However, unlike in the Fediverse, a relay does \emph{not} communicate with other relays; it interacts solely with users. Further, users can push posts to any (including multiple) relays, thereby decoupling the fixed association between a user and an individual server.
Thus, even if individual relays censors content (or simply fail), the posts remain live.
Note, this may also offer greater censorship resistance against other forms of censorship (\eg state-level firewalls), but Nostr does not explicitly strive to address such scenarios.
To ensure integrity, each user is uniquely identified by a public key, and each message is digitally signed. 
These unique features have garnered significant attention from users, and fostered numerous applications built atop of Nostr. To date, the Nostr ecosystem encompasses more than 600 active relays \cite{online-relay}, more than \blue{9 million users (with a profile and contact list)}, and 100 million posts~\cite{nostr-band}.

As the Nostr ecosystem continues to grow, we argue that it is crucial to better understand its characteristics and the associated challenges.
We are particularly curious about whether Nostr encounters similar resilience challenges to that of the Fediverse, such as poor relay (instance) availability, post availability, and the natural tendency towards centralization \cite{IMC19-Mastodon}. 
Critically, we wish to test how effectively Nostr achieves the goal of censorship resistance, and if this comes at the cost of high overhead.
For example, the replication of posts may significantly increase storage overhead and relay-to-user communication costs. To date, these challenges are not understood and we lack any deterministic solutions.

With this in mind, we present the first large-scale study of the Nostr ecosystem. We assemble a dataset of 17.8 million posts, 1.5 million unique pubkeys (users), and 712 relays, spanning the time period from \texttt{2023-07-01} to \texttt{2023-12-31}.
Using this dataset, we focus on two key aspects.

First, we study Nostr from the perspective of \textbf{relays}, inspecting their availability, reliability, and the degree of decentralization. Through the relay-based paradigm, we show that Nostr has accomplished significantly greater decentralization compared to prior Fediverse applications (\S\ref{subsec:relays-distribution}, \S\ref{subsec:relays-hosting}). The users and posts are highly distributed across the relays, attaining significant censorship resilience. Most (93\%) posts can be found in multiple relays, with 178 relays (25\%) individually hosting a significant proportion (more than 5\%) of global posts. The relays are distributed over 44 countries and 151 autonomous systems (ASes). Thus, no single country or autonomous system hosts more than 25\% of the relays, and the majority have only 1 or 2 relays each, making censorship attacks far more difficult.
However, we find that the unstable availability of relays is still a challenge (\S\ref{subsec:relays-availability}). 
We show that 20\% of the relays experience downtime for more than 40\% of the measurement period.
Further investigation suggests that financial sustainability may be a key reason for this, especially for free-to-use relays (\S\ref{subsec:relays-income}). Although Nostr offers support for user donation, our estimates shows 95\% of the free-to-use relays cannot cover their operational cost from this alone.

We next focus on the replication and availability of \textbf{posts}.
A unique innovation of Nostr is the ability to push posts across multiple relays. 
This, however, imposes a trade-off between improved post availability (\ie censorship resistance) and lower replication overhead.
We show that post availability on Nostr significantly surpasses that of Fediverse applications: even when the top 30 relays with the highest number of users or posts are down, over 90\% of the posts remain accessible (\S\ref{subsec:posts-availability}).
Initially, this suggests that Nostr has accomplished its core anti-censorship and resilience goal. 
However, the overhead of post replications is large. We find 616M post replications for 17.8M posts.
This means, on average, a post is replicated across 34.6 relays (\S\ref{subsec:posts-replications}, \S\ref{subsec:posts-redundancy}). Another challenge stemming from post replications is that, due to the Nostr protocol design, in many cases, users must retrieve the same post multiple times from different relays. 
Indeed, our evaluation indicates that 98.2\% of these retrievals are redundant, leading to extremely wasteful consumption of bandwidth ~(\S\ref{subsec:posts-retrieval}).

These results motivate us to study if it is possible to reduce the overhead of post replications. Consequently, we propose a new feature in post publications for Nostr clients (\S\ref{subsec:innovation-reduce}) that automatically limits the number of replications and strategically selects relays for these replications. Our evaluation demonstrates that this feature can effectively reduce 380 million (61.7\%) post replications while preserving the same level of post availability and censorship resistance \blue{under the simulated scenarios of relay-level and AS-level failures}.
Moreover, we propose another feature in post retrieval for Nostr clients (\S\ref{subsec:innovation-optimistic}) to minimize wasted bandwidth by optimistically querying only a subset of known relays. According to our evaluation, this feature effectively decreases wasteful retrievals by 93\%, while ensuring that 99.4\% of the posts can be retrieved.
This paper makes three key contributions:
\begin{enumerate}[noitemsep,topsep=2pt,parsep=0pt,partopsep=0pt, leftmargin=*]
    \item We perform the first large-scale systems study of the Nostr ecosystem, covering 17.8 million posts, 616 million post replications, 1.5 million users, and 712 relays.
    
    \item We confirm that Nostr achieves high content availability, making censorship extremely difficult. However, while we highlight that Nostr is more decentralized with improved post availability and censorship resistance, we show that it comes with significant overhead.
    
    \item  In response, we propose two design improvements: One aims to control the number of post replicas, and one aims to minimize the overhead during post retrieval. We show that \blue{in our data-driven simulation scenarios,} they are effective without causing negative impact on the censorship resistance. 
    
\end{enumerate}

\vspace{-1.5ex}
\section{Nostr Primer}
\label{sec:back}
\vspace{-0.5ex}

\pb{Nostr Overview.}
Nostr is a decentralized social networking system built on a series of open protocols. From a user’s perspective, it can be seen as a micro-blogging service similar to Twitter/X or Mastodon.
However, the underlying architecture is very different.
Nostr is composed of two key elements: \emph{relays} and \emph{clients}. A relay operates as a server, receiving messages from certain users and forwarding them to others (upon request) in an asynchronous manner. Users interact with relays, sending and receiving messages through a client. This client can be any application (mobile, desktop, web) that implements the Nostr client protocol.

Any individual can operate a relay, akin to the Fediverse where anyone can set up an instance. Similarly, end users can select any client implementation, or even build their own compatible client.
For context, Figure \ref{fig:nostr_primer} in the Appendix offers the screenshots of two Nostr clients.
When using these clients, a user can create a post, which is then pushed to the relays of their choice. Similarly, the clients can connect to relays to download posts. Further details about this process are provided in the following paragraphs. 
The design aims to achieve strong censorship resistance by distributing post replications across multiple relays. This makes it difficult for attacks or failures to affect all relays simultaneously. As long as at least one relay continues to host the post, it remains available, thereby ensuring censorship resistance.

\pb{User Account.}
To begin, the user needs an account, which is a public key independent from any relay.
Nostr utilizes asymmetric encryption, identifying each user by their respective public key. 
The user generates a key pair and provides it to the client to facilitate interactions with relays. Crucially, as long as the user maintains their private key, they have ultimate ownership of the account, unlike platforms like Twitter/X or Mastodon where the account can be manipulated or deleted by system admins. This flexibility also enables the user to seamlessly switch between relays while maintaining the same account. This makes it notably different to prior platforms, which retain ultimate account ownership.

\pb{Event.}
The \emph{event} is defined as a specific JSON schema, which serves as the fundamental data structure for data transmission in Nostr.
Nostr then defines various kinds of events for different purposes, such as user profiles (kind-0), posts (kind-1), following lists \ie lists of other users followed by the user (kind-3). A post (kind-1) is the core and most common event, functioning similarly to a tweet on Twitter/X. Together, these events facilitate the operation of a social network.
More details can be found in the protocol specification \cite{nostr-nip01}. 

\pb{Relay Selection.}
To publish and retrieve within the Nostr ecosystem, the user needs to select a set of relays to use. For this, the client maintains a relay list which is configurable by the user. The client connects to all relays in this list for publishing and retrieving posts. Users can find relays to add to the list through three main methods:
\one~the client comes with a default list of relays;
\two~the client identifies relays from a relay index website (\eg \texttt{nostr.watch});
or
\three~other users recommend relays within the user profile or through a specific Nostr event (kind-10002).

\pb{Connecting to Relays.}
Nostr uses WebSockets \cite{websocket} to connect to relays. The Nostr protocol dictates that clients should initiate a single WebSocket connection to each relay and employ it for all forms of communication with this relay.

\pb{Sending to Relays.}
To send an event to relays, the user crafts the content, then uses the client to create and sign the event with their private key before dispatching it to the relays. 
The user can publish events to multiple relays of their choice. The relays store the event and, when requested, forward it to other users. Importantly, the event's authenticity is verified through its signature, eliminating the need to trust the relay.

\pb{Receiving from Relays.}
To receive posts from a relay, the client must create a \emph{subscription} and send it to the relays. 
Each subscription includes a filter that specifies the kinds of events the client wishes to receive.
The relay then forwards the desired events to the client. For instance, if user $F$ follows user $A$, $F$'s client will create a subscription with a filter for $A$'s recent events.
Thus, client $F$ will then connect to the relay and upload the filter.
The relays will then forward $A$'s events to the client and continue to forward new events from $A$ once they are received.

\pb{Zaps.}
Zaps function as monetary tips that one user can send to another, which is also used to fund relays. A user can opt to link their account with a Bitcoin wallet address, thereby enabling them to send or receive Bitcoins in the form of zaps. A successful zap transaction results in the publication of a zap receipt within the relays as a kind-9735 event. More details can be found in \cite{nostr-nip57}.

\vspace{-1ex}
\section{Dataset}
\label{sec:dataset}
\vspace{-0.5ex}

We focus on two key aspects of the Nostr ecosystem: relays and post replications across relays. 
To measure these, we gather two datasets. The first is the Nostr relay dataset, which includes general information, deployment, and availability data of the relays. The second is the event dataset, which contains the events in the relays. This dataset helps us gauge the scale and popularity of the relays, as well as measure post replications and availability. See \S\ref{subsec:ethics} for an ethics discussion.

\vspace{-1.5ex}
\subsection{Relay Dataset}
\label{subsec:dataset-relay}
\vspace{-0.5ex}

\pb{Relay Selection.}
We begin by compiling a list of relay URLs and IP addresses from the \texttt{nostr.watch} website which is widely utilized within the Nostr ecosystem. It maintains a comprehensive index of relays around the world by implementing Nostr's relay discovery and monitoring protocol \cite{nostr-nip66}. We extract the relays from this site on a daily basis from \texttt{2023-07-01} to \texttt{2023-12-31}.
We collect the list of 911 relays that have been online at least once during this period. We successfully retrieve event data from 712 of them.  We cannot obtain event data from the remaining 199 relays due to three reasons: \one some relays (58\%) have no events published, indicating that they are either not actively used or solely set up for testing purposes; \two certain relays (41\%) remain online for only a short period before going constantly offline; and \three some relays (1\%) require payment for accessing their data. Consequently, all analyses conducted in this study exclude these relays from which we cannot obtain event data.

\pb{Relay Information.}
Relays provide server metadata (a relay information document) to clients. This document provides clients with information on server capabilities, administrative contacts, and various server attributes, such as
content limitations, community preference, and payment information. 
We provide an example relay information document in Appendix \ref{appendix:sample_nip11}, and more details can be found in the protocol specification \cite{nostr-nip11}.
This information also allows us to separate the relays into free relays and paid relays. In the case of a paid relay, users are required to pay a certain amount of bitcoin. While most paid relays only charge for posting to the relay, a small number charge for reading (we exclude them because we cannot crawl them). Of the 712 relays in our dataset, 625 are free and 87 are paid.

\pb{Autonomous Systems.}
We then query \texttt{IP-API} \cite{IP-API} with the IP address of the relay to get the AS hosting the relay, as well as the geographical location (country) of the relay server.

\pb{Relay Availability.}
To monitor the availability of relays, we utilize the feature provided by \texttt{nostr.watch}. The \texttt{nostr.watch} website conducts frequent periodic measurement to test the availability of the relays. 
Every 15 minutes, we take a snapshot of the availability status of relays reported by \texttt{nostr.watch}. Our monitoring period is from \texttt{2023-10-01} to \texttt{2023-12-31}. This allows us to map the exact downtime of each relay.

\vspace{-1.5ex}
\subsection{Event Dataset}
\label{subsec:dataset-event}
\vspace{-0.5ex}

\pb{Event Collection.}
We connect to each relay in our dataset to gather \blue{the} posts, users and zap related events. We divide the collection process into several parallelized jobs and distributes the workload evenly across the relays, preventing any single relay from being overwhelmed.

\pb{Posts.}
As described in \S\ref{sec:back}, a post is a kind-1 event in Nostr, which is roughly equivalent to a tweet on Twitter/X.
However, in contrast to Twitter/X, we emphasize that posts can be  replicated across multiple relays (but is identified by a unique ID).
Thus, we regard each unique ID as an individual post and record the relays in which it is replicated.
We crawl 712 relays for the time period from \texttt{2023-07-01} to \texttt{2023-12-31}, and get a total of 17,851,871 Posts.

\pb{Users.}
Each unique pubkey represents an individual user. We follow this convention throughout the remainder of the paper.
We identify active users who have posted at least once between \texttt{2023-07-01} and \texttt{2023-12-31} from our posts dataset. We obtain 1,558,891 unique pubkeys in total. 
Subsequently, we construct the user dataset by querying the relays for user profiles (kind-0 event) and the user's following list (kind-3 event) using the obtained public keys.

\pb{Zaps.}
To investigate the financial incentives of operating a relay, we collect kind-9735 events associated with the account of relay operators, which is used for the zap transactions as described in \S\ref{sec:back}. 
We crawl the relays for the time period from \texttt{2023-07-01} to \texttt{2023-12-31}, and get a total of 252,905 zap transactions, with the amount summed up to be 305,427,302 Satoshi (1 Bitcoin = 100 million Satoshi). During the measurement period, it is roughly 137,000 USD.

\pb{Other Events.}
Note, there are other kinds of events serving different purposes. We do not to collect these, as some kinds are only visible to specific users (such as direct messages or group chats) and others are specified by more advanced extensions of the Nostr protocol, and not all relays have implemented these features. Since our measurement focuses on the overall Nostr ecosystem, we exclude these specific event kinds from our analysis.

\vspace{-1.5ex}
\section{Characterizing Nostr Relays}
\label{sec:relays}
\vspace{-0.5ex}

Nostr is designed to achieve censorship resistance by distributing post replicas across multiple relays, making it difficult for attacks or failures to occur simultaneously on all relays. Thus, in this section, we focus on the properties of the Nostr relays. There are several crucial points must be considered to ensure effective censorship resistance:
\one Are the posts distributed to a sufficiently large number of relays to remain robust?
\two How are these posts distributed across ASes and countries, and is this robust to single points of failure?
\three How reliable are the collection of relays, hosting each individual post?
\four What is the cost of relay operation, and is it sustainable?

\vspace{-1.5ex}
\subsection{Distribution of Posts \& Users}
\label{subsec:relays-distribution}
\vspace{-0.5ex}

\begin{wrapfigure}{r}{0.5\linewidth}
    \centering
    \vspace{-6ex}
    \includegraphics[width=\linewidth]{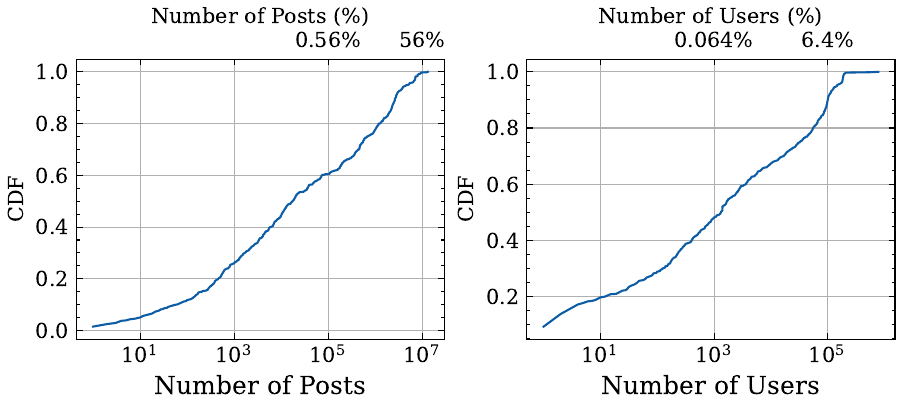}
    \vspace{-5ex}
    \caption{(a) Distribution \& CDF of the number of posts in a relay; (b) Distribution \& CDF of the number of users in a relay.}
    \label{fig:rq1_popularity}
    \vspace{-3ex}
\end{wrapfigure}

We first evaluate the degree of decentralization within Nostr at the relay level. This gives an idea of how susceptible posts are to specific takedown attacks.
To do this, we inspect the distribution of posts and users across the relays, \ie the number of posts in the relay and the number of users using the relay.
We define a user is using a relay if they have at least one post in the relay. 
Previous research has highlighted user-centralization and content-centralization problems in the Fediverse \cite{IMC19-Mastodon}, \ie a small number of instances host the vast majority of posts and users. While we aim to determine if these findings also apply to Nostr, we also note a vital difference: in Nostr, a user and a post can exist in multiple relays. We posit this uniqueness may result in distinctive characteristics.

\pb{Distribution of Posts \& Users.}
Figure \ref{fig:rq1_popularity}a illustrates the distribution and cumulative distribution function (CDF) of the number of posts in a relay. We observe that the top relay hosts 73\% of the posts. 
At first glance, this may suggest a highly centralized system. However, it is important to consider that a post can exist in multiple relays.
Taking this into account, the result reveals that, although 178 relays (25\%) host a significant proportion (more than 5.6\%) of the posts, the sum of these proportions is far larger than 100\% --- \blue{on average, each post is replicated 34.6 times}.
This confirms that posts are extensively replicated across multiple relays. This could be viewed as a decentralized backup of posts, serving to mitigate the risk of failure or censorship in a single relay. Therefore, the result stands as a favorable signal of decentralization. 
We also note that this result can be biased if there is a heavy-tailed distribution, where a few posts are heavily replicated while most are not. 
\blue{However, we later confirm that this is not the case in \S\ref{subsec:posts-replications}, where the results are plotted in Figure \ref{fig:rq2_post_replication_hist}}.

Figure \ref{fig:rq1_popularity}b illustrates the distribution of the number of users in a relay. We observe a similar distribution to the distribution of posts. 
This suggests that users are actively utilizing multiple relays for publishing (and likely retrieving) posts. Once again, this highlights Nostr's distinctive feature of detaching users from specific servers, mitigating the risk of censorship.

Overall, the distribution patterns of posts and users demonstrates the effective decentralization at the relay level. Unlike the Fediverse, we do not find evidence for excessive user-centralization or post-centralization at the relay level in Nostr.

\begin{figure}[]
    \centering
    \includegraphics[width=0.36\linewidth]{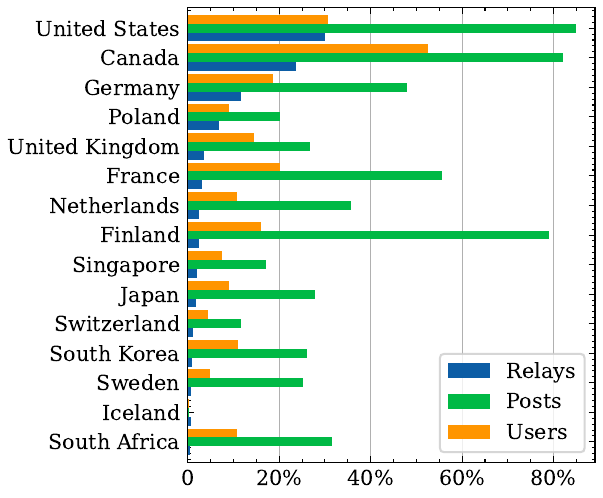}
    \includegraphics[width=0.45\linewidth]{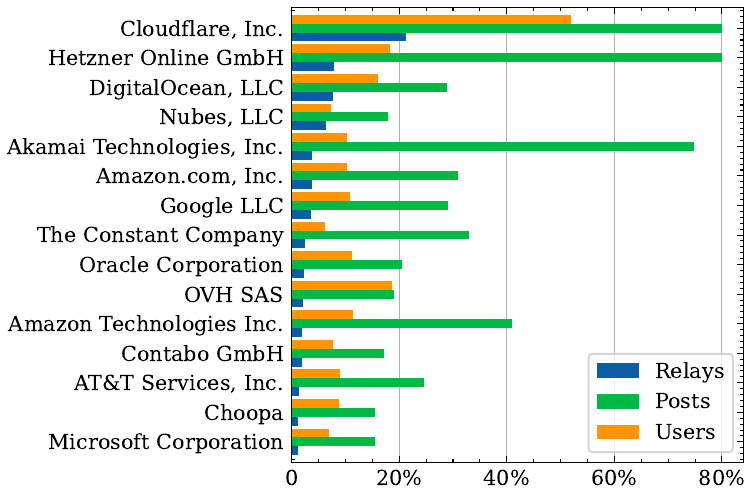}
    \vspace{-2ex}
    \caption{Percentage of relays, posts, and users in the top 15 regions and ASes, ranked by number of relays.}
    \label{fig:rq1_region_bar}
    \vspace{-3.5ex}
\end{figure}

\vspace{-1.5ex}
\subsection{Relay Hosting}
\label{subsec:relays-hosting}
\vspace{-0.5ex}

\begin{wrapfigure}{r}{0.5\linewidth}
    \centering
    \vspace{-8.5ex}
    \includegraphics[width=0.49\linewidth]{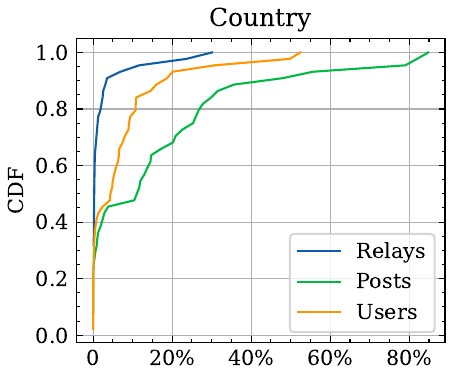}
    \includegraphics[width=0.49\linewidth]{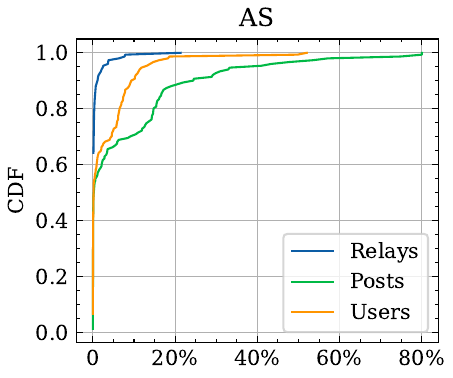}
    \vspace{-5ex}
    \caption{CDF of the percentage of relays, posts, and users in (a) the countries; (b) the ASes. The x-axis begins with a near-zero number but not 0, since relays of no posts are already excluded as described in \S\ref{subsec:dataset-relay}.}
    \label{fig:rq1_region_AS_cdf}
    \vspace{-2.5ex}
\end{wrapfigure}

We next evaluate the degree of decentralization of Nostr at the country and autonomous systems (AS) level. Spreading posts across multiple country and AS level jurisdictions is vital for delivering on Nostr's censorship resistance goals, to prevent all administrators within a jurisdiction being forced to remove content.
Like instances in the Fediverse, relays in Nostr follow a bottom-up approach, where administrators have the autonomy to choose where to host their relays. 
Prior work revealed that there is infrastructure-centralization in the Fediverse \cite{IMC19-Mastodon}. That is, the hosting of instances and posts are centralized in a few countries and Autonomous Systems (ASes).
Thus, we hypothesize that Nostr may encounter similar challenges, whereby the majority of relays fall under the remit of a small number of countries or ASes, making them vulnerable to regional or network-level failure and censorship attacks.
To test this, we investigate the hosting of relays in Nostr to see if relays, posts or users are overly centralized in certain locations.

\pb{Countries.}
Figure \ref{fig:rq1_region_AS_cdf}a displays the CDF of the percentage of relays, hosted posts, and users in different countries.
Additionally, Figure \ref{fig:rq1_region_bar}a illustrates the percentage of relays, hosted posts, and users for the top 15 countries ranked by the number of relays.
In stark contrast to the Fediverse, we observe that the distribution of relays is quite decentralized, with around 52\% of countries having fewer than two relays. 
That said, upon closer inspection, we see that users and posts are concentrated in the top countries, such as the US, which hosts 85\% of posts and 30\% of users. 
Nonetheless, it is crucial to recall that, unlike Fediverse instances, a post and a user can belong to multiple relays, which creates backups that will persist even if some relays censor or fail.
As a result, most of the top 15 countries host at least 20\% of the posts, and Canada, France, Germany, and Finland host more than 40\%.
This distribution is actually a positive indicator of decentralization, as it suggests that a significant portion of Nostr posts and users are spread across different countries. 
\blue{This outcome appears to be more favorable compared to Mastodon, where a prior study revealed that 89.1\% of all toots were concentrated on instances in Japan, the US, and France \cite{IMC19-Mastodon}.}

\pb{ASes.}
Figure \ref{fig:rq1_region_AS_cdf}b displays the CDF of the percentage of relays, posts, and users available via different ASes.\footnote{We note that some ASes listed could be Content Delivery Networks (CDNs), who do not directly host the posts themselves. We are unable to identify the specific servers behind these CDNs as currently there is no reliable approach to do so.} 
Additionally, Figure \ref{fig:rq1_region_bar}b illustrates the percentage of relays, posts, and users for the top 15 ASes ranked by the number of relays.
We observe a similar trend to countries, as the distribution of relays across ASes is also surprisingly decentralized: 64\% of ASes host only one relay. 
As shown in Figure \ref{fig:rq1_region_bar}b, among the top 15 ASes, all of them host more than 15\% of the posts. Once again, this serves as a positive indicator of decentralization and offers greater protection to AS-level censorship attacks, whereby a given AS could shutdown entire sets of relays.

Note, this distribution also contrasts significantly with the findings in the Fediverse. \blue{According to \cite{IMC19-Mastodon}, toots and users in Mastodon were primarily centralized around just three major ASes, as administrators were naturally drawn to well-known and cost-effective cloud providers.} One possible explanation for this disparity is that Nostr relays are typically more lightweight than Fediverse instances, requiring fewer computational resources. Consequently, administrators may find better pricing options from smaller providers rather than relying on major providers like AWS.

\vspace{-1.5ex}
\subsection{Relay Availability}
\label{subsec:relays-availability}
\vspace{-0.5ex}

Recall, the main design purpose behind Nostr's greater decentralization is superior availability and resilience to censorship. 
Thus, we next explore the availability properties of the relays.
In this context, availability refers to the ability to establish a connection with the relay. Similar to Fediverse instances, we posit that the voluntary nature of certain relay operations might lead to unique availability characteristics.

\begin{figure}
    \centering
    \includegraphics[width=0.245\textwidth]{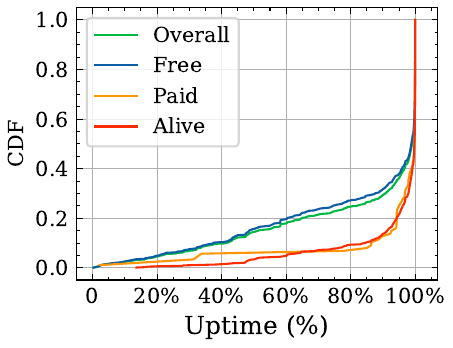}
    \includegraphics[width=0.49\textwidth]{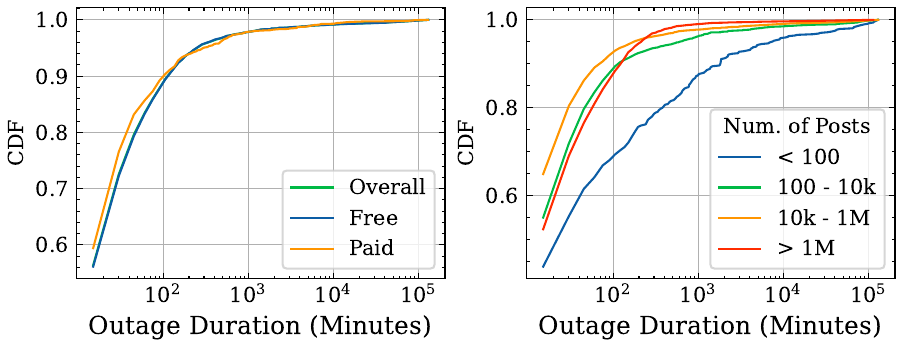}
    \includegraphics[width=0.245\textwidth]{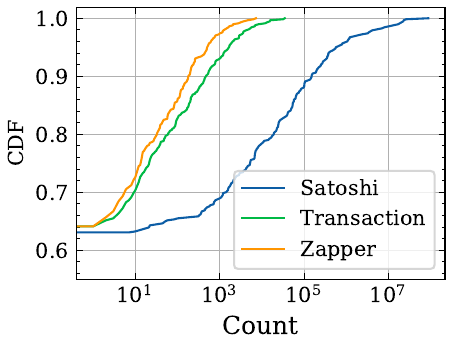}
    \vspace{-2ex}
    \caption{(a) CDF of the percentage of \bblue{uptime} for free, paid, alive, and all relays; (b) CDF of the outage duration for free, paid, and all relays; (c) CDF of the outage duration for relays with different number of posts; (d) CDF of the count of zapped Satoshi (amount of zap), zap transactions and zappers for relays.}
    \label{fig:rq1_down_duration_cdf}
\end{figure}

\pb{Relay Availability.}
Based on the relay dataset described in \S\ref{sec:dataset}, we calculate the \bblue{uptime} of relays from \texttt{2023-10-01} to \texttt{2023-12-31}. The CDF of the \bblue{uptime} percentage is presented in Figure \ref{fig:rq1_down_duration_cdf}a.
From the figure, we observe that the availability of some relays is relatively good, with 50\% of the relays being available for more than 99\% of the specified time period. However, there are also some relays with extremely poor availability. Approximately 20\% of the relays experience downtime for more than 40\% of the time. 
However, further investigation reveals that most of these relays are offline for extended periods and do not even reactivate by the end of our measurements. We therefore define relays that have been offline for more than a week and remain so at the end of our measurements as \emph{dead}. In total, we identify 132 dead relays. 
When removing these dead relays from the analysis, as shown by the red line in Figure \ref{fig:rq1_down_duration_cdf}a, the availability of improves significantly, with 90\% of the relays being operational for more than 80\% of the time.

Overall, this finding confirms that relay failures are relatively common in Nostr. This broadly aligns with that observed in the Fediverse~\cite{IMC19-Mastodon} (although the most significant failures can be attributed to dead relays). However, fortunately, in Nostr, users have the flexibility to switch between relays easily. As a result, the impact of relay unavailability in Nostr may not be as severe as in instance unavailability in Fediverse.

\pb{Outage Duration.}
We next investigate the outage duration. 
For each outage, we compute its duration and plot the CDF in Figure \ref{fig:rq1_down_duration_cdf}b and \ref{fig:rq1_down_duration_cdf}c .
We do not exclude dead relays in this analysis, as dead relays can be considered as relays with a long outage duration.
First, we compare free and paid relays to determine if paid relays also exhibit better performance in terms of shorter outage duration.
Interestingly, we find that the outage durations for paid and free relays are nearly identical, as depicted in Figure \ref{fig:rq1_down_duration_cdf}b. 
The majority of outages (56\%) last for less than 15 minutes. However, there is also a long tail, with 10\% of the outages lasting for more than 100 minutes.

Next, we group the relays based on the number of posts to observe if the scale of the relays correlates with the outage duration. As shown in Figure \ref{fig:rq1_down_duration_cdf}c, we notice that larger relays perform better, particularly in the tail end. For the group of relays with over 1 million posts, only 1\% of the outages last for more than 1000 minutes, whereas the corresponding figure for the group of relays with less than 100 posts is 88\%. 
This finding suggests that although the downtime may be similar, the outage durations for larger relays are shorter. This implies that while the availability of larger relays may not be consistently stable, it is unlikely for them to experience prolonged outages lasting several hours. 
\bblue{On the other hand, relays that remain operational more consistently are therefore more likely to have posts successfully pushed to them.
Thus, they tend to accumulate more posts.}

\subsection{Relay Incomes}
\label{subsec:relays-income}

One of the challenges within the Fediverse is the cost of running instances, coupled with a lack of incentives for doing so \cite{Allen2023-ALLEOD}.
This is also one reason why some instances have low availability, with some even being abandoned and subsequently closed. 
This issue also applies to Nostr, as the lack of financial income may lead to a poor availability of certain relays and the system may become centralized around well-resourced individuals (who do not require financial support).
Therefore, we examine the possible incomes of the relays, and importantly, whether the income generated is sufficient to cover operational costs, and thus to keep the decentralized operation of the Nostr ecosystem.
We argue that this is key to relay availability and therefore censorship resistance, as a lack of money could result in relays shutting down.

\pb{Paid Relays.}
One direct source of revenue is the admission fee for paid relays. This fee is a one-time cost for users to publish to the relay. 
We find that the admission fee is relatively low for some of the relays, with 35\% charging less than 1k Satoshi (approximately \$0.45 USD). We argue that this minimal fee makes it challenging to sustain the relay's operation. This implies that, for these relays, the primary purpose of the admission fee is to act as a barrier, allowing only genuine users while keeping spammers out, rather than generating profits. 
On the other hand, there are also relays that impose a higher admission fee, with 15\% of relays charging more than 10k Satoshi (approximately \$45 USD). This higher fee may potentially help in covering the expenses associated with operating the relay. We explore this in the subsequent paragraphs.

\pb{Zaps.}
Another available revenue stream is through zaps sent by users to relay administrators. To estimate this, we analyze the zap transactions to the public keys specified in relay information. If no public key is provided in the relay information, we consider the zap income as zero. 
Figure \ref{fig:rq1_down_duration_cdf}d shows the CDF of the total zap amount (note that 52\% of the transactions do not disclose the zap amount publicly), count of zap transactions, and zappers. 
We find that the majority (64\%) of relays do not receive any zaps.
Among the remaining relays, some of them do receive a relatively high number of zap transactions, with the 90th percentile at 471. However, the zap amount itself is not substantial, with the 90th percentile receiving a total of 150k Satoshi (approximately 67 USD).

\begin{wrapfigure}{r}{0.45\textwidth}
    \centering
    \includegraphics[width=\linewidth]{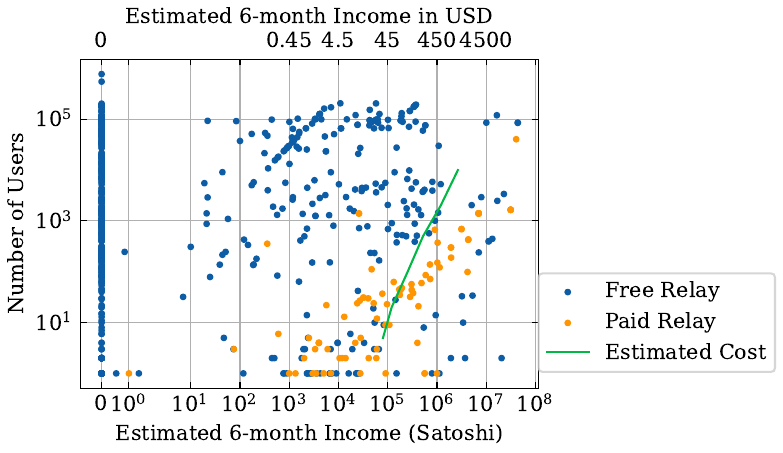}
    \vspace{-5ex}
    \caption{Number of users \vs the estimated 6-month income of free and paid relays. The green line is the estimated 6-month cost of running a relay.}
    \label{fig:rq1_relay_income_2}
    \vspace{-3ex}
\end{wrapfigure}

\pb{Enough to Cover the Cost?}
We finally assess whether the income generated is sufficient to cover operational costs.
For this, we estimate the 6-month income (\texttt{2023-07-01} to \texttt{2023-12-31}) of relays. 
To achieve this, we first calculate the zap amount for the relays. Then, for paid relays, we add the product of the admission fee and the number of active users. For relays operational for less than 6 months, we calculate the daily average income and multiply it by 184 to estimate the income for the 6-month period.
Figure \ref{fig:rq1_relay_income_2} illustrates the estimated income \vs the number of users as a representation of the relay's scale (See Appendix \ref{apendix:relay_income} for figures with different Bitcoin prices). The green line roughly estimates the cost of operating a relay based on the cost of running a Mastodon instance of similar size, as listed in \cite{mastodon-host-price}.

It is evident that, barring a few outliers, most of the free relays with more than 10 users are unable to cover their operating costs. 
However, most paid relays of medium scale (with 50 -- 1k users) are able to cover their operational costs.
This constitutes a key threat to long-term viability.
It is therefore necessary for relays to obtain additional revenue (often donated by the voluntary administrator). 
The trend shown in the figure is significant in terms of magnitude. Therefore, the conclusion remains valid, even though there may be some errors due to transactions not publicly disclosing the zap amount and the accuracy of estimating the operational cost based on Mastodon.

This result suggests that, akin to the Fediverse, most Nostr relays are operated by enthusiasts without clear financial incentives. The number of dead relays and the lower availability of free relays revealed in \S\ref{subsec:relays-availability} may be a consequence of this situation. 
\blue{However, the observed difference in uptime between paid and free relays (\S\ref{subsec:relays-availability}) is not as stark as this financial disparity might suggest. 
This implies that non-monetary factors, such as hobbyist enthusiasm and the relative operational simplicity of the Nostr software, could also be drivers of reliability.
Consequently, while financial unsustainability is a threat to the long-term existence of free relays, our current data does not support a strong causal link between a relay's payment model and its day-to-day uptime.}

\summary{
\one Nostr achieves greater decentralization than previous Fediverse applications, with posts replicated across multiple relays in 44 countries and 151 autonomous systems, confirming high censorship resistance.
\two Unstable availability of relays is still a challenge for Nostr, akin to Fediverse applications. 
\three Financial sustainability \blue{might be a complicating} factor in relay availability issues, as most Nostr relays are operated by enthusiasts without clear financial incentives to support the operational cost.}
\section{Post Availability \& Replication}
\label{sec:posts}

In  \S\ref{subsec:relays-availability}, we find that certain relays (including the larger ones hosting over 1 million posts) experience periods of downtime.
However, one of the key design innovations of Nostr is that the posts can be replicated across multiple relays, so that the availability of posts will not be significantly impacted by censorship  or failures on a few relays. 
Indeed, in \S\ref{subsec:relays-distribution} and \S\ref{subsec:relays-hosting}, we find many posts are replicated across multiple relays, ASes, and countries. 
Thus, we next evaluate the effectiveness of this replication design innovation in enhancing post availability and censorship resistance, alongside the overhead associated.

\vspace{-1.5ex}
\subsection{Number of Post Replications}
\label{subsec:posts-replications}
\vspace{-0.5ex}

\begin{wrapfigure}{r}{0.5\textwidth}
    \centering
    \includegraphics[width=\linewidth]{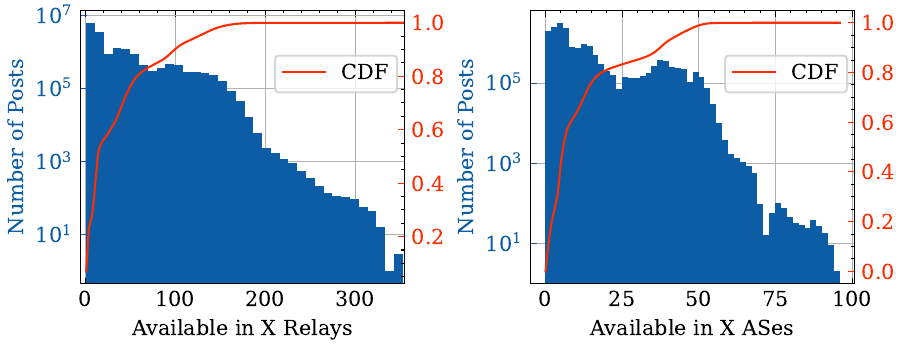}
    \vspace{-5ex}
    \caption{(a) Distribution and CDF of the number of relays where a post is available; (b) Distribution and CDF of the number of ASes where a post is available.}
    \label{fig:rq2_post_replication_hist}
    \vspace{-2ex}
\end{wrapfigure}

We first assess the replication of posts, at a per-post level. 
Figure \ref{fig:rq2_post_replication_hist} presents the CDF of the number of relays and ASes in which a post is replicated. 
We see remarkable levels of replication: the majority of posts are replicated across multiple relays and ASes, with the average being 34.6 and 12.4, respectively.
While the high volume of post replications contributes to increased post availability and censorship-resistance, it also leads to an excess of redundancies. This may incur unnecessary costs, \eg relay storage space. We delve deeper into this aspect in the subsequent subsections.

\vspace{-1.5ex}
\subsection{Post Availability under Attack}
\label{subsec:posts-availability}
\vspace{-0.5ex}

To evaluate the robustness of posts to censorship attacks, we simulate relay removals using our dataset. 
Failures could be triggered by censorship attacks from third parties (\eg DDoS) or simply a shutdown by the administrator.
We sequentially remove the most pivotal relays and ASes, and then measure the number of posts that remain available.
We employ two ways of ranking the most pivotal relays and ASes to remove: measured by the largest number of \one~posts and \two~users. We perform an additional ranking for ASes based on the number of relays. Note, we define a post as available if it is replicated on at least one relay.

Figure \ref{fig:rq2_remove_top_N}a shows the percentage of available posts after the removal of the top $X$ relays.  In stark contrast to the Fediverse \cite{IMC19-Mastodon}, Nostr's availability and censorship resistance remains impressively robust, hovering around 90\% even after the top 50 relays, ranked by the number of users, are removed. 
Even when removing the top relays based on the number of posts, the availability remains above 90\% only until the top 30th relay is removed (after which it experiences a sharp decline, culminating at 71\% at the top 50th).
That said, we argue that the availability is still fairly impressive.
Figure \ref{fig:rq2_remove_top_N}b also displays the proportion of accessible posts following the elimination of the top X ASes. 
Despite the failure of entire ASes, over 80\% of posts remaining available even after the removal of the top 10 ASes. To provide a comparison, the post availability plunges to less than 10\% after the removal of the top 10 ASes hosting the instances \cite{IMC19-Mastodon}. This is largely driven by the more uniform distribution of relays across distinct ASes, making them more resilient to individual AS takedowns.
These findings confirm significant censorship resistance within the Nostr ecosystem, a result largely attributable to the extensive replication of posts.

\vspace{-1.5ex}
\subsection{Post Redundancy}
\label{subsec:posts-redundancy}
\vspace{-0.5ex}
Despite the impressive post availability attained by Nostr, we suspect that this comes at the cost of excessive overhead. This involves an ``over-replication'' of some posts across numerous (unnecessary) relays. For example, our data shows that, 50\% of posts are replicated across over 14 relays, and 25\% are replicated across more than 49 relays. Arguably, this level of redundancy is unnecessary, except in the face of catastrophic wide area attacks or failures. Consequently, we next measure the possible optimization space of post replications using our dataset.

To do this, we investigate the number of replications that can be eliminated if we cap the maximum replication of a post at $N$. Specifically, for each post $P$ with $M_P$ replications, where $M_P > N$, we calculate $\Delta_P = M_P - N$. Subsequently, we compute the result by summing up $\Delta_P$ for all posts $P$.
This result represents the optimization space under the assumption that exceeding $N$ replications is unnecessary.

\begin{figure}[]
    \centering
    \includegraphics[width=0.64\textwidth]{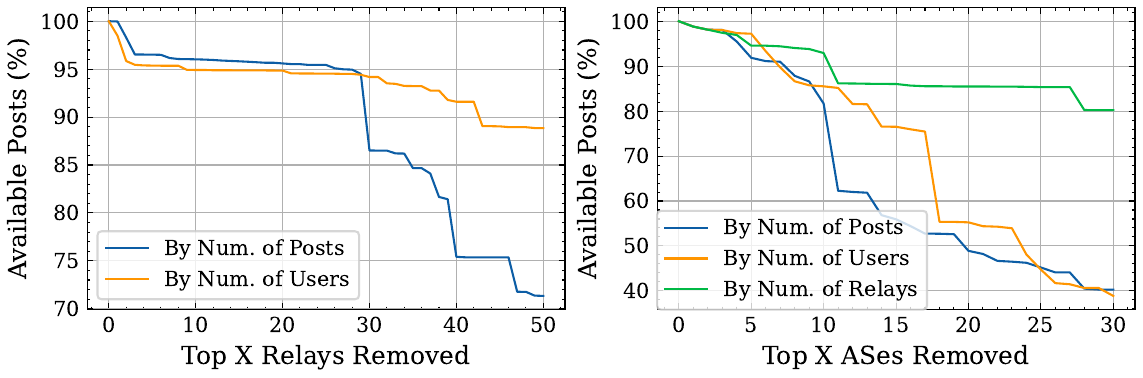}
    \includegraphics[width=0.32\textwidth]{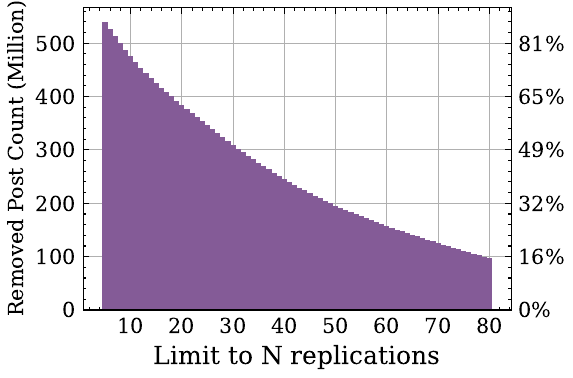}
    \vspace{-2.5ex}
    \caption{(a) Percentage of available posts when top X relays are removed; (b) Percentage of available posts when top X ASes are removed; (c) Number of removed post replications when the maximum number of post replications is limited to N.}
    \label{fig:rq2_remove_top_N}
    \vspace{-3.5ex}
\end{figure}

The results are depicted in Figure \ref{fig:rq2_remove_top_N}c.
By restricting the maximum replication to 20, it is possible to eliminate 380 million (61.7\%) post replications. Further reducing this limit to 10 leads to the removal of an additional 100 million post replications, totaling 480 million (77.9\%). This substantial reduction suggests that, if each user only posts to a maximum of 10-20 relays, over 400 million client-relay communications for posting could be saved, along with the storage space required for the over 400 million replications. We later revisit the implications that this has on resilience to relay failures or censorship takedowns.

\vspace{-1.5ex}
\subsection{Duplicated Retrieval From Relays}
\label{subsec:posts-retrieval}
\vspace{-0.5ex}

The prior subsection has highlighted that Nostr's approach to post replication and substantially improves availability and censorship resistance. However, this comes at the cost of storage overhead. In this subsection, we explore another unexpected overhead created by this approach: A client often must redundantly retrieve the same post from several relays, resulting in unnecessary bandwidth usage. 
This occurs because in Nostr's design, a client will maintain connections to all relays the user is using, and retrieve posts from all these relays. 
For example, if user $A$ follows user $B$, and both are using relays $R_1$, $R_2$, and $R_3$, then $B$ will send a post $P$ to all three relays. Consequently, $A$ will receive the identical post $P$ from $R_1$, $R_2$, and $R_3$ respectively.
This is done because a client does not necessarily know which relay will be hosting the next event of interest.
However, the ideal scenario would be for $A$ to retrieve $P$ from just one of the three relays. 
To date, Nostr has no solution to elegantly deal with this duplicated retrieval.

\pb{Estimating Traffic Volume.}
%
We begin by estimating the amount of wasted traffic caused by the duplicated retrieval of posts. 
Here, we assume that a user will retrieve all the posts from the users they are following via all the relays they are using, as defined by the follower relationships.
First, we construct a directed graph $G$ of all users to represent following relationships. In this graph, a vertex $V$ signifies a user, and a directed edge $(U, V)$ indicates that user $U$ follows user $V$.

\pb{Retrieval Count of a Post over an Edge.}
For a post, $P$, that is shared over a following edge $(U, V)$, we calculate the number of times it is retrieved across relays. 
We refer to this as the \emph{retrieval count}.
First, we get $S_{P}^{R}$, representing the set of relays where the post $P$ is duplicated.
Second, we get $S_{U}^{R}$, representing the set of relays utilized by user $U$. 
Last, we compute $|S_{P}^{R} \cap S_{U}^{R}|$, which corresponds to the retrieval count of $P$.

\pb{Traffic over an Edge.}
For a following edge $(U, V)$, we calculate the volume of traffic over it. 
First, we get $S_{V}^{P}$, which represents all posts published by user $V$.
Note, these are equal to all posts transmitted over edge $(U, V)$.
Second, for every post in $S_{V}^{P}$, we calculate its retrieval count. 
We then calculate the \emph{retrieval size}, by multiplying the retrieval count by the size of the post.
Last, we sum the retrieval counts and retrieval traffic for all posts within $S_{V}^{P}$ to obtain the total retrieval count and retrieval size over the edge $(U, V)$.

\pb{Traffic over the Following Graph.}
We also calculate the volume of traffic over the entire follower graph $G$.
For every edge in $G$, we compute the traffic over the edge, and sum up the results.

\begin{figure*}
    \centering
    \includegraphics[width=\textwidth]{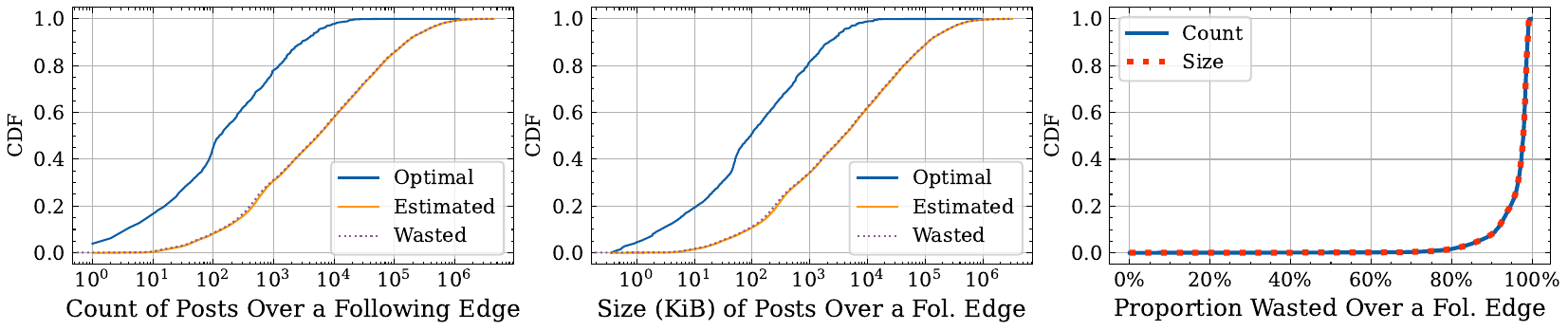}
    \vspace{-4.5ex}
    \caption{CDF of (a) the retrieval count of posts transmitted over a following edge; (b) the retrieval size of posts transmitted over a following edge; \blue{(c) the percentage of wasted retrieval count and retrieval size of posts transmitted over a following edge.}}
    \label{fig:rq3_edge_traffics}
    \vspace{-2.5ex}
\end{figure*}

\pb{Results.}
%
We use the above method to estimate the quantified amount of traffic.
In total, the follower graph $G$ contains 3.5M follower relationships between users.
The CDF of the retrieval count and retrieval size of posts over an edge is illustrated in Figure \ref{fig:rq3_edge_traffics}a and \ref{fig:rq3_edge_traffics}b, respectively.
For comparison, we also plot the ``optimal'' case.
This assumes that each post is retrieved only once (\ie from a single relay).
We calculate the wasted volume of traffic by 
subtracting the optimal value from the estimated value. We refer to this as the \emph{wasted} volume, which is also plotted in the figures.

From Figure \ref{fig:rq3_edge_traffics}a and \ref{fig:rq3_edge_traffics}b, we observe that the estimated download traffic exchanged is significantly higher than the optimal value (where each post on an edge is retrieved only once).
As a result, the estimated amount of traffic wastage is almost identical to the estimated volume of traffic, \ie nearly all download traffic in Nostr is redundant. 
On average, the optimal retrieval count for an edge is 1130 posts, while the actual estimated value is 62,839 posts. This is a significant cost to pay for the additional censorship protections.

To more clearly depict this, Figure \ref{fig:rq3_edge_traffics}c plots the CDF of the percentage of wasted traffic per follower edge. This graph shows a significant volume of wasted traffic. 
We observe that 98\% of the traffic is wasted for over 50\% of the following edges. 
We find that 98.2\% of the download traffic in Nostr is wasted, totaling 144 TiB
.
This is because the majority of clients are connected to multiple relays, thereby redundantly downloading the same objects many times.

Overall, we find that the duplicated retrievals from relays results in substantial traffic waste.
This leads to unnecessary costs in operating the relays and can have a negative impact on the Nostr ecosystem. It also has a detrimental impact on clients with limited resources. 
We argue that certain simple innovations could mitigate this issue; thus, we explore possible solutions in \S\ref{subsec:innovation-optimistic}.

\summary{\sloppy \one Post availability on Nostr significantly surpasses that of Fediverse applications, making censorship difficult without a concerted large-scale attack upon many relays.
\two However, the overhead of post replications is large. On average, a post is replicated across 34.6 relays, introducing significant storage overhead. 
\three Due to the Nostr protocol and extensive post replications, users retrieve the same post multiple times from different relays, leading to significant wastage of bandwidth for relays and clients. }

\section{Potential Nostr System Improvements}
\label{sec:innovation}

In \S\ref{sec:posts}, we found that the replication of posts in Nostr leads to significant overhead. 
Thus, we next investigate potential improvement that can reduce the associated overhead.
The goal is to provide references and insights for the community to further improve the operation of Nostr, while retaining the same level of censorship resistance.

\vspace{-1.5ex}
\subsection{Improving Storage Redundancy}
\label{subsec:innovation-reduce}
\vspace{-0.5ex}

\S\ref{subsec:posts-redundancy} shows that, by restricting the maximum number of replications, it is possible to eliminate millions of replications. 
Thus, we propose an enhancement for Nostr clients, where they set a cap of $N$ relays to publish their posts to.
This, however, may impact post availability and censorship resistance. 
\bblue{We are aware that reducing this redundancy, even slightly, creates a trade-off between theoretical availability and practical efficiency. Our position is that the current level of replication is excessive, imposing a severe burden on relay operators for a level of resilience that may not be necessary in practice. Our goal is to find a more pragmatic balance that ensures the network's long-term health.}

Thus, it is important that a client intelligently selects the most resilient composition of $N$ relays.
We strive to make the relay selection mechanism flexible. The strategy for selecting relays and the number of $N$ is therefore configurable by the user to suit their specific needs. Yet, intuitively, we would want a post to be replicated across different ASes to prevent central points of failure. 
Thus, we propose an AS-based strategy that prioritizes availability, aiming to distribute the post across the maximum possible number of ASes.
We evaluate our proposed enhancement by simulating the situation where we limit the maximum number of replicas to $N$ and then compare the post availability to the original using the same method as in \S\ref{subsec:posts-availability}.

\pb{Evaluation Method.}
To assess post availability after limiting replicas to a maximum of $N$, we simulate new scenarios with different $N$ values, derived from our existing dataset. Within the original dataset, each post, $P$, is associated with a map $P \rightarrow S$, where $S$ represents the set of relays where $P$ is replicated. 
To reflect the scenario where the maximum number of replicas is limited to $N$, this mapping is then modified. For any post $P$ with replicas exceeding $N$ (\ie $|S| > N$), we create a subset $S'$ from $S$ such that $|S'| = N$, and let $P \rightarrow S'$. 
To evaluate the trade-off of $N$ and the availability, we employ the same approach as detailed in Section \S\ref{subsec:posts-availability} (\ie simulating relay failures) to measure the post availability within these new scenarios. 
We run the experiments in 4 scenarios with $N=$ 5, 10, 15, and 20, respectively. 
Additionally, we include the scenario where $N=2$ to illustrate how availability diminishes when the replication number is extremely low.

We experiment with two strategies for selecting $S'$: random (as a baseline) and our proposed AS-based strategy.
Specifically, with the random strategy, for every $P \rightarrow S$ with  $|S| > N$, we construct subset $S'$ by randomly selecting $N$ relays from $S$, and let $P \rightarrow S'$.
For the the AS-based strategy, for every $P \rightarrow S$ with  $|S| > N$, we construct subset $S'$ such that the relays in $S'$ span as many ASes as possible, and let $P \rightarrow S'$. 
To do this, we start by selecting one random relay from $S$ for each possible AS. If we do not have enough, we randomly select additional relays from $S$ until the total number equals N. This is designed to be inherently simple and not require coordination across multiple clients or relays.

\begin{figure}[]
    \centering
    \includegraphics[width=0.48\linewidth]{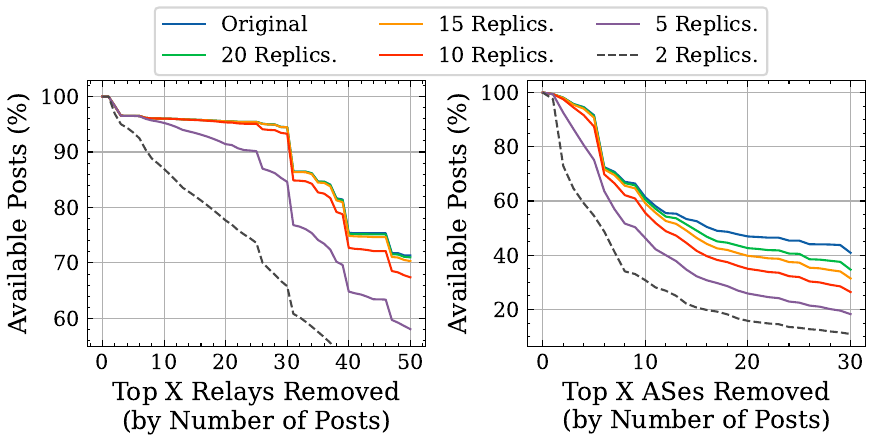}
    \vline
    \includegraphics[width=0.48\linewidth]{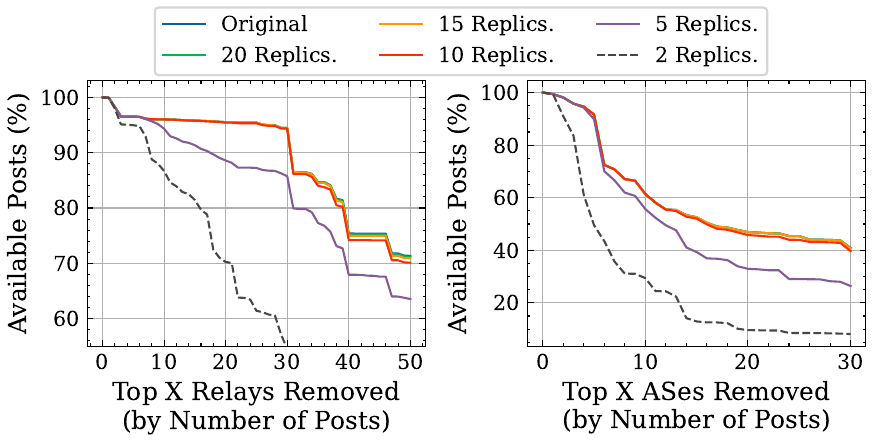}
    \vspace{-2.5ex}
    \caption{Percentage of available posts when top X relays are removed and when top X ASes are removed, if the maximum number of post replications is (a) reduced to N randomly; (b) reduced to N based on AS.}
    \label{fig:rq2_reduce_to_N_random}
    \vspace{-2.5ex}
\end{figure}

\pb{Evaluation Results.}
We use the same method as in \S\ref{subsec:posts-availability} to simulate relay failures and thus compare the post availability under attack to the original.
Figure \ref{fig:rq2_reduce_to_N_random} shows the results of the impact on availability when removing either \one~individual relays; or \two~entire ASes (and therefore all relays hosted within). 
To quantify the difference in availability between the new scenario and the original, we begin by defining $A(N, X)$ as the availability when the maximum replication is restricted to $N$, and the top $X$ relays are removed. 
This corresponds to each point depicted in the figure. Subsequently, to capture the difference in availability between the given scenario and the original, we introduce $\Delta^A(N, X) = A(N, X) - A(O, X)$, where $O$ denotes the original scenario.
In order to encapsulate the overall difference between a line in the figure and the original, we define $D^A_{relay}(N, X) = \Sigma^{X}_{i=1} \Delta^A(N, i) / X$. This captures the average difference in availability across all $X$ values.
Similar to $D^A_{relay}$, we also define $D^A_{AS}(N, X)$ for the removal of $X$ ASes.

Figure \ref{fig:rq2_reduce_to_N_random}b plots the results for the AS-based strategy.
Even when removing several relays or ASes, if the maximum number of post replicas is limited to $N \ge 10$, the post availability (percentage of available posts) is very close to the original. That is, throughout the measured $X$ values, $D^A_{relay}(N, X) < 1\%$ and $D^A_{AS}(N, X) <1\%$ for $N \ge 10$.
Additionally, this is also significantly better than the random strategy, as depicted in Figure \ref{fig:rq2_reduce_to_N_random}a, especially in the scenario where several ASes (and all relays hosted within) are removed. To quantify this, we find $D^A_{relay}(10, 50)$ is -0.39\%, and $D^A_{AS}(10, 30)$ is -0.53\%. In contrast, for random relay selection, $D^A_{relay}(10, 50)$ is -1.15\%, and $D^A_{AS}(10, 30)$ is -8.38\%. \bblue{This outcome is anticipated, as this relay selection strategy directly opposes scenarios where the entire AS is removed.}

\pb{Evaluation Summary.}
The results confirm that our proposed solution is effective.
For reasonable values of $N$ (such as 15 and 20), we dramatically reduce the post redundancy \blue{while maintaining high levels of post availability and censorship resistance against our simulated relay-level and AS-level failures.}  Further, by choosing our proposed AS strategy, it is possible to choose smaller values of $N$ such as 10. In the case where the number of replications is restricted to $N = 20$, 380 million (61.7\%) post replications can be eliminated.

\blue{However, it is important to note that this AS-spanning replication optimization could introduce a load imbalance, particularly for relays that are the sole one hosted in their AS.
This could open up potential threats.
For example, an adversary could exploit this by identifying and targeting these unique relays in small ASes to disproportionately impact the overall network. Our current resilience evaluation assumes whole AS-level failures and does not model this more surgical targeted relay attack. 
Therefore, Nostr's resilience under more targeted adversarial strategies remains an open question.}

\subsection{Improving Retrieval Traffic Waste}
\label{subsec:innovation-optimistic}

The issue of duplicated retrieval of posts by clients (\S\ref{subsec:posts-retrieval}) arises from the fact that, in the vanilla Nostr implementation, clients retrieve the posts of every user they follow from all relays in their relay list. 
This creates substantial communications overhead for both clients and relays. An intuitive solution might be for a client to send a filter requesting the relay to exclude any prior posts they already possess. However, in practice, such a solution is impractical. This is because clients connect in parallel to all relays, to enable real-time communications. Sending a filter would require a client to connect to each relay sequentially one-by-one (so that the filter can be updated). This could result in a delay of several minutes for the client to refresh the timeline, which many users may find unacceptable. It also introduces additional complexity and network load (\eg sending a filter also cost additional bandwidth) on both the client and relay-side, which Nostr strives to avoid.

Thus, we propose a design improvement that avoids any additional client-relay communication, where the client only retrieves posts from only a small subset of relays (rather than from all relays in their list). 
This, however, demands the intelligent selection of which relays to contact.
To select which $N$ relays to retrieve posts of user $V$ from, we first rank all relays based on the number of user $V$'s posts stored there. We then select the top $N$ relays that host the most posts by user $V$. 
We posit that this lightweight strategy increases the likelihood of accessing most posts. 
However, even with the best relays, posts can be missed. Thus, to evaluate this trade-off, we test our proposed design by simulating the scenario where clients only retrieve posts from a selected subset of $N$ relays.

\begin{wrapfigure}{r}{0.6\textwidth}
    \centering
    \includegraphics[width=\linewidth]{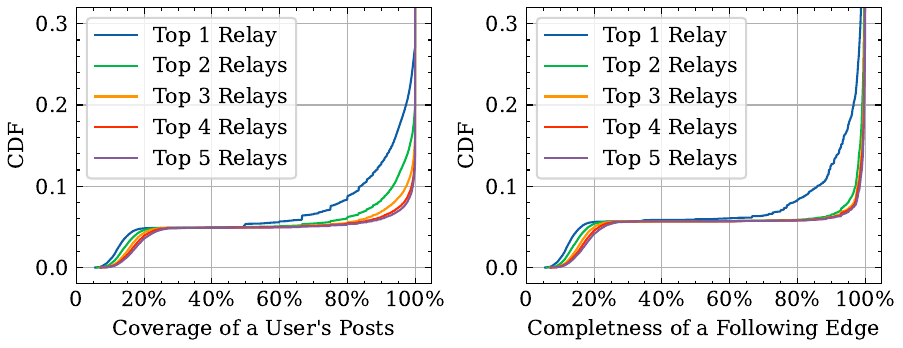}
    \vspace{-4ex}
    \caption{CDF of (a) post coverage for users; (b) completeness of following edges with optimistic retrieval.}
    \label{fig:rq3_optimistic_coverage_cdf}
    \vspace{-2ex}
\end{wrapfigure}

\pb{Evaluation of Post Coverage.}
We first evaluate the strategy from the post sender's perspective, where we examine the proportion of posts that our approach of selecting the $N$ relays can cover for a sender. 
We denote $S_{U}^{P}$ as the set of all posts of user (sender) $U$. 
We denote $S_{R}^{P}$ as the set of all posts contained within the relay $R$. 
The intersection, $S_{U}^{P} \cap S_{R}^{P}$ represents the posts made by sender $U$ that are stored by relay $R$. 
Then, we define the coverage of $R$ for $U$ as $({|S_{U}^{P} \cap S_{R}^{P}|}\ /\ {|S_{U}^{P}|}) \times 100\%$, \ie the percentage of the posts of $U$ that is covered by $R$. 
Similarly, for a set of relays $S^R$, we first compute the set of all posts contained within the relays in $S^R$, as $S_{S^R}^{P} = \bigcup_{R \in S^R} S_{R}^{P}$.
Then the coverage of $S^R$ for $U$ is $({|S_{S^R}^{P} \cap S_{R}^{P}|}\ /\ {|S_{U}^{P}|}) \times 100\%$.

For each user (sender) $V$, we evaluate the coverage of the selected $N$ relays (the top $N$ relays ranked by the number of $V$'s posts) for $V$.
We construct 5 sets with the $N=$ 1, 2, 3, 4, and 5, respectively.
Figure \ref{fig:rq3_optimistic_coverage_cdf}a presents the CDF of the coverage offered by the five set sizes of relays for each sender. 
We see that, by retrieving solely from the top four or top five relays, 90\% of senders can ensure that all recipients see their posts (\ie 100\% coverage).
This confirms that by strategically opting for a small number of relays to push posts to, the post coverage for the majority of senders is satisfactory.

\begin{wraptable}{r}{0.44\textwidth}
    \scriptsize
    \vspace{-2ex}
    \centering
    \begin{tabular}{r|c|cc|cc}
    \hline\hline
        \multicolumn{1}{c}{} & \multicolumn{1}{c}{comple-} & \multicolumn{2}{c}{Wasted Count} & \multicolumn{2}{c}{Wasted Size}  \\
        {} & {teness} & Wasted &  \% & Wasted & \% \\
    \hline
        Original & - & 218.3B & - & 143.9T & - \\
        $N = 2$ & 99.0\% & 3.8B & 1.7\% & 2.5T & 1.7\% \\
        $N = 3$ & 99.2\% & 7.7B & 3.5\% & 5.1T & 3.5\% \\
        $N = 4$ & 99.3\% & 11.6B & 5.3\% & 7.7T & 5.4\% \\
        $N = 5$ & 99.4\% & 15.4B & 7.0\% & 10.2T & 7.1\% \\
    \hline\hline
    \end{tabular}
    \caption{The post completeness and potentially wasted count and size of posts over the following graph, when only retrieving from the selected $N$ relays, with N = 2, 3, 4, and 5.}
    \label{tab:optimistic_retrieval}
    \vspace{-8ex}
\end{wraptable}

\pb{Evaluation of Completeness of Following Edges.} \\
We next evaluate our design from the perspective of users who want to retrieve the post.
To look at which posts each retriever should retrieve, we turn to the follower graph. 
This tells us which posts each user should retrieve. 
We construct the follower graph $G$, and evaluate the post retrieval completeness over $G$. We evaluate the completeness of each following edge $(U, V)$. Specifically, let $S^P_V$ denote the set of all posts by $V$, and $O^P_V$ denote the set of $V$'s posts that can be retrieved by $U$ from the selected $N$ relays for $V$. 
The \emph{completeness} is then computed as $({|O^P_V|}\ /\ {|S^P_V|}) \times 100\%$. Figure \ref{fig:rq3_optimistic_coverage_cdf}b shows the CDF of the completeness across the follower edges. Again, we observe that nearly 90\% of following edges achieve 100\% coverage.

\pb{Saved Traffic.}
Finally, we evaluate the traffic savings attained by reducing the level of redundant retrievals. To do this, we use the same approach as described in \S\ref{subsec:posts-retrieval}, while assuming clients only retrieve posts of a user from their selected set of $N$ relays.
Specifically, in the original method in \S\ref{subsec:posts-retrieval}, for a following edge $(U, V)$, we retrieve from $S_U^R$. This represents the set of relays utilized by user $U$. 
We then use $S_{V_N}^R$, which represents the selected $N$ relays for user $V$ (top $N$ relays with largest number of $V$'s posts). 
Table \ref{tab:optimistic_retrieval} presents the results for $N=$ 2, 3, 4, and 5. 
The table shows that even when $N$ is set to 5, it is possible decrease the duplicated retrieval count by 93\%, while still achieving 99.4\% of post completeness.

\pb{Evaluation Summary.}
The results highlight that our proposed design, \blue{under our simulated scenarios,} substantially reduces wasted traffic, without heavily compromising the completeness of the retrieved posts. Although a small ($<$1\%) number of posts are missed, we posit that this is acceptable in most practical scenarios as the Nostr protocol, by its design, does not guarantee completeness.
Considering that numerous relays are lightweight and subject to the costs associated with metered bandwidth, we contend that our proposed design significantly enhances the efficiency and overall sustainability of the Nostr ecosystem, without introducing substantial new complexity.


\vspace{-1.5ex}
\section{Related Work}
\vspace{-0.5ex}

To the best of our knowledge, this research is the first large-scale analysis of the Nostr ecosystem, and Nostr is the first decentralized social networks at scale to prioritize censorship resistance as its primary objective. However, there has been ongoing research on censorship resistance techniques and decentralized social networks.

\pb{Censorship Resistance Techniques.}
Numerous techniques combat Internet censorship. Proxy-based systems, for instance, use millions of short-lived proxies \cite{10.1007/978-3-642-31680-7_13}, while TLS-based tools utilize encrypted channels to hinder content monitoring \cite{frolov2019use}. The Tor network provides anonymity by routing traffic through volunteer-operated servers, concealing user activity \cite{tor1, tor2}. Video steganography conceals information within multimedia files for covert communication \cite{10.1145/3488932.3517419, Barradas2020}. Core network-based circumvention includes methods like decoy routing \cite{karlin2011decoy} and domain fronting \cite{fifield2015blocking}. These varied techniques underscore the multifaceted approach to combating censorship. We complement this prior work with our focus on Nostr, a censorship-resistant microblogging service.

\pb{Decentralized Social Networks.}
Early efforts in decentralized social networks utilized peer-to-peer (P2P) technologies, with examples like Safebook \cite{cutillo2009safebook} and PeerSoN \cite{buchegger2009peerson}. Recently, federated networks such as Mastodon and PeerTube, forming the Fediverse \cite{mansoux2020seven}, have gained prominence, typically using protocols like W3C ActivityPub \cite{activitypub} for server-to-server communication. Research on these federated systems has investigated Mastodon's centralization tendencies \cite{IMC19-Mastodon}, growth patterns \cite{cava2021understanding}, user behaviors \cite{cava2022information, cava2022network}, and content moderation challenges in services like Pleroma \cite{Conext21-Pleroma, Sigmetrics22-Toxicity}. More recently, Bluesky has gained attention, while many of its components are still in a (semi-)centralized mode \cite{balduf2025bootstrappingsocialnetworkslessons, 10.1145/3646547.3688407}.
Nostr presents a different architecture. Unlike Fediverse applications where servers interoperate, Nostr relays are independent. Users connect directly to multiple relays, enhancing censorship resistance and account portability, as accounts are not tied to a specific relay. This allows users to replicate posts across various relays for better availability.

\bblue{Other decentralized social networks offer different approaches to this design space. Peer-to-peer (P2P) systems like Secure Scuttlebutt (SSB) \cite{SSB} use a gossip protocol where users replicate the feeds of those they follow. This shifts the storage burden to the user's device and can lead to significant client-side overhead and synchronization challenges. Similarly, various blockchain-based social networks (e.g., memo.cash \cite{memo}, Steemit \cite{guidi2020socioeconomic}) embed social data directly onto a blockchain \cite{arquam2021, guidi2020, jiang2019}. While this offers strong immutability, it incurs extreme overhead in terms of storage, transaction fees, and network bandwidth, which must be borne by the entire network.
Nostr, while using public-key cryptography for post signing, is distinct: it neither employs blockchain technology (though users can use cryptocurrency for payments), nor complex P2P data replication and validation like SSB. Instead, Nostr prioritizes a simpler, lightweight, and user-friendly design. That said, our findings indicate that Nostr also experiences network overhead, suggesting it is not exempt from the classic resilience-overhead tradeoff inherent in censorship-resistant social networks.}

\vspace{-2ex}
\section{Conclusion, Limitations and Future Work}
\vspace{-0.5ex}

\pb{Concluding Remarks.}
This paper presents a comprehensive study of the Nostr ecosystem, an emerging censorship-resistant decentralized social network. 
Nostr represents a new design philosophy for building decentralized social networks: it is intentionally simple and offers minimal features, while prioritizing key attributes that many people expect from a decentralized platform: openness, high availability, and censorship resistance. Overall, we highlight the operational characteristics of such a system in contrast to traditional Fediverse applications, revealing its strengths, limitations, and trade-offs where we also propose potential improvements. We hope these insights can aid in the development of other decentralized social networks, especially for those aiming to prioritize censorship resistance.
We wish to flag several limitations, which form the basis of our future work.

\pb{\bblue{Measurement Method Limitations.}}
\blue{The exclusion of offline or paywalled relays from our dataset might skew the results to reflect the behavior of publicly accessible relays.
Our findings about relay availability rely on third-party data from \texttt{nostr.watch}. Although this tool is widely used within the Nostr community, conducting independent probes to validate the results would enhance the reliability of our claims regarding relay availability. 
That said, this approach also offers the advantage of ensuring that the availability results are not influenced by factors such as being banned by relays.
The analysis of geolocation and IP-to-AS mapping depends on the \texttt{IP-API} service, which may introduce minor mappuing errors. Thus, future research could improve geolocation accuracy by utilizing multiple sources for geolocation data.}
\bblue{Our current estimation of relay operating costs is based on Mastodon instances. To enhance accuracy, future research could improve this by conducting surveys with relay operators.}
\bblue{When measuring the availability under attack, we only consider attacks in relay and AS level. However, attacks can occur not only at the AS-level but also at the country level. Studies have examined several instances of country-level censorship, for example, in China \cite{298156} and in Turkmenistan \cite{10.1145/3543507.3583189}. Future research may consider these broader levels of attack, which would provide a more comprehensive understanding of availability challenges in different geopolitical contexts.}

\pb{\bblue{Proposed Potential Improvements.}}
\bblue{Our proposed optimizations are intentionally designed to modify only the client side, a choice made to avoid complicated coordination and ensure easy adoption. A limitation of this client-centric approach is that our evaluation, while demonstrating bandwidth savings, does not measure the potential impact on user-perceived latency or content consistency during query fallbacks. Furthermore, our simulation assumes uniform adoption and does not account for adaptive or alternative user behaviors, such as circumventing replication caps or strategically selecting relay subsets, which could alter real-world performance.}

\bblue{At a higher level, our proposed optimizations, while effective at reducing operational overhead, are not a complete solution to the problem of relay sustainability caused by the lack of incentives. Our work squarely addresses the cost side of the economic equation for operators but does not introduce new revenue streams. Therefore, a critical direction for future work is to explore and evaluate systemic solutions, such as micropayments or tokenization, that provide clear financial incentives to ensure the ecosystem's long-term health.}

\pb{\bblue{Longitudinal Dataset Limitations.}}
\bblue{Our dataset captures the Nostr network during a six-month window ending in late 2023. Given Nostr's rapid evolution, our findings represent a specific point in time, and the network's characteristics regarding user base, relay stability, and client behavior may have since changed. This highlights critical avenues for future research. A continuous, longitudinal analysis with more recent data is essential to track long-term trends and validate the continued relevance of our conclusions. Further, new decentralized social networks like Bluesky \cite{balduf2025bootstrappingsocialnetworkslessons, 10.1145/3646547.3688407} have emerged recently and attracted a significant number of users. Conducting a comparative performance analysis against these networks would certainly be interesting.}

\vspace{-1.5ex}
\subsection*{Acknowledgments}
\vspace{-0.5ex}
This work was supported in part by the Guangzhou Science and Technology Bureau (2024A03J0684), Guangdong provincial project 2023QN10X048, the Guangzhou Municipal Key Laboratory on Future Networked Systems (2024A03J0623), the Guangdong Provincial Key Lab of Integrated Communication, Sensing and Computation for Ubiquitous Internet of Things (No.2023B1212010007), the Guangzhou Municipal Science and Technology Project (2023A03J0011), Guangdong provincial project (2023ZT10X009), and the 111 Center (No. D25008).

\bibliographystyle{ACM-Reference-Format}
\bibliography{sample-base}

\newpage
\appendix

\section{Appendix}

\subsection{Ethical Considerations}
\label{subsec:ethics}

The relay dataset only comprises public infrastructure information and data that has been voluntarily disclosed by the relay operator. The event dataset includes both user and post information that are publicly available to anyone. To prevent any mishandling of data, we adhere to established ethical procedures for social media data \cite{townsend2016social, townsend2017ethics} and only collect publicly available data. Additionally, we do not analyze specific users or posts, nor do we try to link users to their real personal identities.

\blue{Our approach to crawling the relays is designed to be minimally intrusive. Each time we connect to a relay, we gather events from a specific time period rather than retrieving all events in the relay. This method is similar to how a typical user refreshes their timeline in a client application (In fact, our crawler is a modified Nostr client). Consequently, we believe this approach imposes minimal load on the relays, ensuring efficient and respectful data collection.}

\subsection{Overview of Nostr}
\begin{figure*}[]
    \centering
    \includegraphics[width=\textwidth]{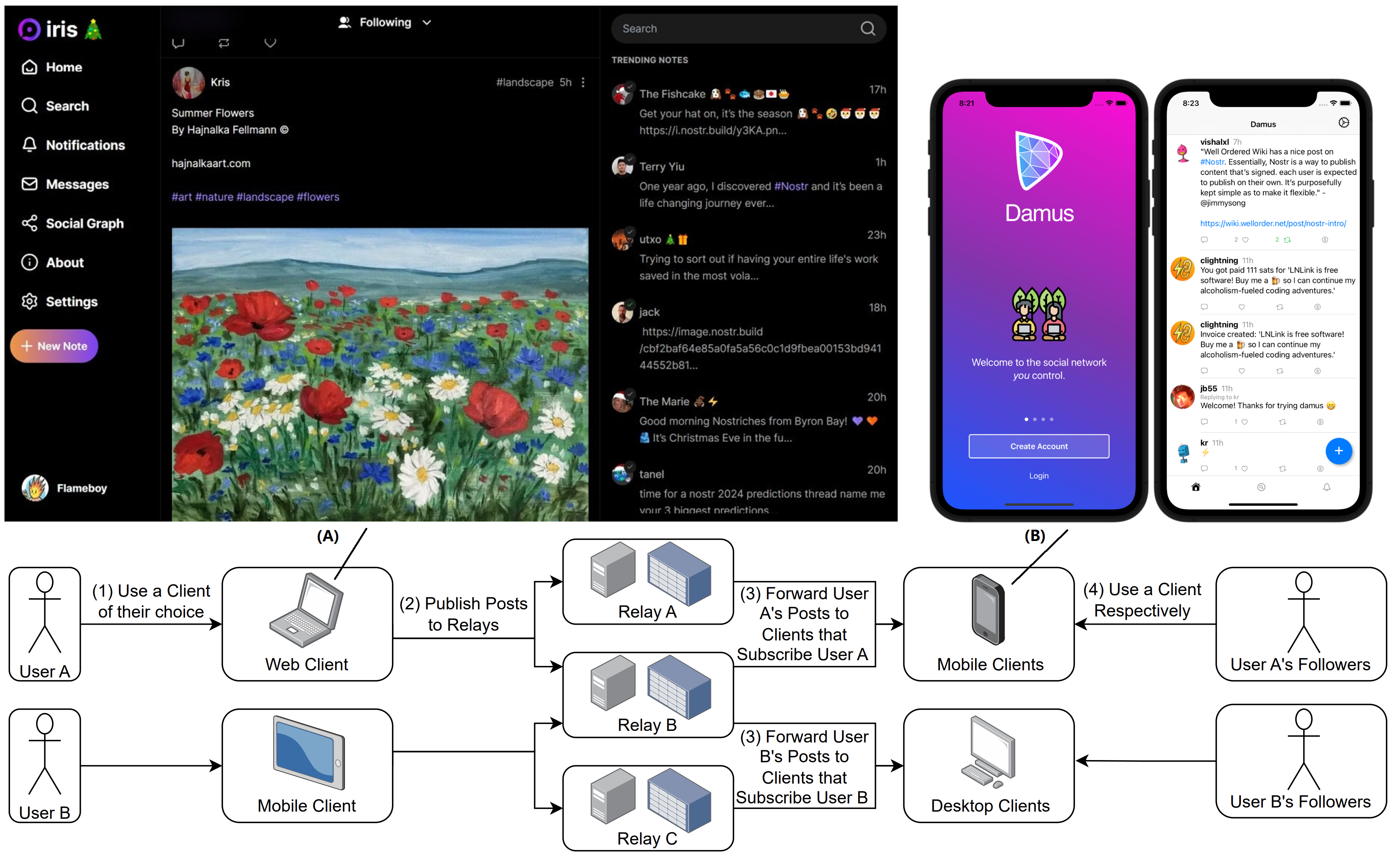}
    \caption{Overview of Nostr. (1) The user uses a client of their choice; (2) Through the client, the user publish the post to relays of their choice; (3) Relays forward the post to clients that have a subscription indicating it wants the message; (4) The client shows the post to the user. (A) Screenshot of Iris, a web client for Nostr. (B) Screenshot of Damus, an IOS client for Nostr.}
    \label{fig:nostr_primer}
    \vspace{5ex}
\end{figure*}

Figure \ref{fig:nostr_primer} offers the screenshots of two Nostr clients.\footnote{Screenshot of Iris is retrieved from \url{https://www.alza.cz/nostr-decentralizovana-sit-pro-svobodnejsi-sdileni-obsahu-na-internetu}, and screenshot of Damus is retrieved from \url{https://github.com/damus-io/damus}.} 

\newpage
\subsection{Sample Relay Information Document}
\label{appendix:sample_nip11}

{\scriptsize \begin{lstlisting}
{
    "contact": "wino@nostr.wine",
    "description": "A paid nostr relay for wine 
        enthusiasts and everyone else.",
    "fees": {
        "admission": [
            {
                "amount": 18888000,
                "unit": "msats"
            }
        ],
        "retention": []
    },
    "limitation": {
        "auth_required": false,
        "max_event_tags": 2000,
        "max_limit": 1000,
        "max_message_length": 131072,
        "max_subid_length": 71,
        "max_subscriptions": 50,
        "min_pow_difficulty": 0,
        "payment_required": true
    },
    "name": "nostr.wine",
    "payments_url": "https://nostr.wine/invoices",
    "pubkey": "4918eb332a41b71ba9a74b1dc64276
        cfff592e55107b93baae38af3520e55975",
    "supported_nips": [1, 2, 4, 9, 11, 12, 15, 16, 
        20, 22, 28, 33],
    "version": "main"
}
\end{lstlisting}}

\subsection{Additional Figures for Relay Income}
\label{apendix:relay_income}

\begin{figure}[h!]
    \centering
    \includegraphics[width=\textwidth]{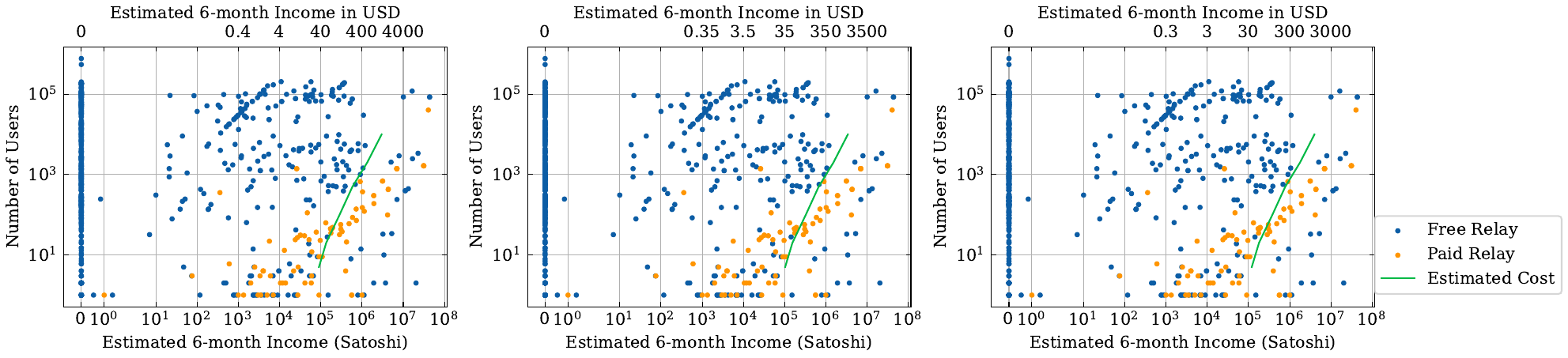}
    \caption{Number of users \vs the estimated 6-month income of free and paid relays. The green line is the estimated 6-month cost of running a relay. The three figures are based on different Bitcoin prices.}
    \label{fig:relay_income_alt}
\end{figure}

Figure \ref{fig:relay_income_alt} shows the number of users \vs the estimated 6-month income of free and paid relays. The green line is the estimated 6-month cost of running a relay. The three figures are based on different Bitcoin prices.

\end{document}